\documentclass[prb,showpacs,amsfonts,amssymb,amsmath,preprint]{revtex4}

\usepackage{times}
\usepackage{graphicx}

\begin{document}

\title{On the Hydrodynamic Boundary Condition for Superfluid Flow}

\author{Yves Pomeau$^{1,2,3}$ and David C. Roberts$^3$}  

\affiliation{$^1$Laboratoire de Physique statistique de l'Ecole normale sup\'erieure, Paris, France \\ $^2$Department of Mathematics, University of Arizona, Tucson, AZ \\ $^3$ Theoretical Division and Center for Nonlinear Studies, Los Alamos National Laboratory, Los Alamos, NM} 


\newsavebox  {\UR}
\sbox{\UR}{{$\ \stackrel{\rightarrow}{u}\ $}}
\newcommand{\urt}{\usebox{\UR}}

\newsavebox{\VR}
\sbox{\VR}{{$\ \stackrel{\rightarrow}{v}\ $}}
\newcommand{\vrt}{\usebox{\VR}}

\newsavebox{\UL}
\sbox{\UL}{{$\ \stackrel{\leftarrow}{u}\ $}}
\newcommand{\ult}{\usebox{\UL}}

\newsavebox{\VL}
\sbox{\VL}{{$\ \stackrel{\leftarrow}{v}\ $}}
\newcommand{\vlt}{\usebox{\VL}}

\newsavebox{\LSIM}
\sbox{\LSIM}{{$\ \stackrel{\textstyle<}{\sim}\ $}}
\newcommand{\lsim}{\usebox{\LSIM}}

\begin{abstract}
We discuss the hydrodynamic boundary condition for a superfluid moving tangentially to a rough surface.  Specifically, we argue that the scattering of quantum fluctuations off surface roughness affects the nature of the boundary condition, and that this has important consequences including a theorized critical speed and the presence of normal fluid at any nonzero speed, even if the boundary is held at zero temperature.  This hydrodynamic boundary condition is relevant not only for superfluid helium experiments but also for experiments with trapped dilute Bose-Einstein condensates, in particular those involving atomic waveguides near surfaces.  
\end{abstract}
 
\maketitle

\section{Introduction}
Much of what is accepted about the nature of superfluid flow and its instability is built on a few concepts laid down by Landau \cite{landau}.  However, many aspects of superfluidity remain mysterious.  One such is the discrepancy between Landau's predicted critical velocity in superfluid Helium and that observed in experiment \cite{vortexshedding}.  

Allen and Misener \cite{allen}, in their letter reporting one of the first observations of superfluidity in liquid Helium, suggested that superfluidity is due to a lack of friction of the fluid with the walls \cite{freeslip}.  Given that the connection between the nature of superfluidity and its boundary conditions was clearly made so early on, it is somewhat surprising how little attention has been paid to superfluid boundary conditions or, more specifically, the boundary conditions of a superflow moving tangentially to a rough wall (see however \cite{schwarz}). 

We believe that these boundary conditions merit closer study.  This paper revisits the broadly assumed boundary conditions on superflow, and hopefully sheds some light on enduring enigmas like the value of the critical velocity in superfluid Helium experiments.  In particular, this paper looks into how the presence of quantum fluctuations affect the boundary conditions on a superfluid, and proposes a resulting effective boundary condition (eq. (\ref{eq:firstattempt})).  The motivation for our view that quantum fluctuations may significantly alter the boundary conditions stems from our previous calculation of a drag force on a localized object; we showed the critical velocity effectively becomes zero for an infinitely extended geometry due to the quantum fluctuations in a superfluid \cite{roberts1, roberts2}.  This discussion should be important for recent experiments directing Bose-condensed atoms through waveguides near a surface (see e.g. \cite{surface,schumm,fortagh}).  It might also provide an explanation for the sensitivity to surface roughness seen in a recent experiment where a vibrating grid is immersed in superfluid Helium \cite{mcclintock}.  Other experimental scenarios involving dilute Bose-Einstein condensates (BECs) and superfluid Helium will be discussed further at the end of this article.

In order for the findings to have broad relevance for current superfluid Helium and dilute BEC experiments, the geometry considered in this paper will be that of the superfluid system contained by rough walls moving at a certain speed $v_s$  relative to the superfluid component.  Wherever it is potentially ambiguous, we will refer to the coherent component as the superfluid component, the remaining fluid (i.e. excitations) as the normal fluid component, and both components of the fluid together as the superfluid, or the superfluid system.  The container is assumed to be a finite system, such as a spinning torus, that locally resembles a long cylinder.  However, the condition of uniformity of the phase in such a geometry does not play a crucial role in the ensuing discussion.  

\section{Landau's two-fluid picture}
Let us begin with Landau's conception of superfluidity.  Landau \cite{landau} wrote `Physically it is obvious that the interaction of the fluid with the moving walls of a tube cannot cause a motion of the whole liquid at once. The motion must begin with the excitation of the inner movements in the layers of liquid close to the walls of the tube, i.e. with the excitation of rotons and phonons in the liquid'.  In his phenomenological picture, excitation of rotons and phonons was the (sole) mechanism through which the walls would interact with the superfluid.  Because such excitations could only be created if the superfluid's motion relative to the wall was above a certain velocity, this led him to postulate the existence of a critical velocity below which the superfluid would flow completely without energy dissipation.  His reflections on superfluid systems implied the following two conditions at the boundary:  
\begin{equation}
\vec{{v}_{n}}=0
\end{equation} 
\begin{equation}
\vec{n} \cdot \vec{{v}_{s}}=0,
\end{equation} where the first is the usual non-slip boundary condition for a viscous fluid, which determines the velocity $\vec{{v}_{n}}$ of the normal fluid component composed of the excitations of the fluid; and the second is the boundary condition of a perfect potential flow, obeyed by the superfluid (or coherent) component with velocity $v_s$.  Khalatnikov later generalized these boundary conditions by including the thermal conductivity of the solid boundary \cite{kha} \cite{heatflux}.  In this paper, however, we will assume that the boundary is perfectly insulating | with zero thermal conductivity | and thus we will not consider heat flux across the boundary.    

Landau's critical-velocity argument involves representing the moving `ground state' of a superfluid flowing tangentially to the container wall as a straightforward Galilean transform of the fluid's true ground state in the non-moving reference frame. Landau makes the implicit assumption that one can ignore the containing walls' roughness when performing this Galilean transform.  For if one took into account the boundary roughness then the ground state boosted to give the fluid a velocity tangential to the wall is not a possible eigenstate of the system.  

To illustrate, consider a superflow moving in the $x$-direction.  In the absence of a containing wall, the fluid's wavefunction would have the property of Galilean invariance. That is, if 
\begin{equation}
\Psi_{0}(\vec{{r}_{1}}, \vec{{r}_{2}}, \vec{{r}_{3}},\ldots)e^{-i\frac{E _{0}t}{\hbar}}
\end{equation}
is a solution of the Schr\"{o}dinger equation with a well defined eigenvalue $E_{0}$ for the energy, where $\vec{{r}_{1}}$,\ldots are the position of all the particles, then 
\begin{equation}
\Psi_{0}(\vec{{r}_{1}} - \vec{v}t, \vec{{r}_{2}}-\vec{v}t, \vec{{r}_{3}}-\vec{v}t,\ldots)e^{-i\frac{E_{0}t}{\hbar}} e^{-i \frac{m{\vec{v}}.\Sigma_{i}\vec{{r}_{i}}}{\hbar}}
\end{equation}
is also a solution, $m$ being the mass of the particles. This remains true in the presence of a flat wall, which can be represented by a potential $U(z)$.  If the wall were rough however, it would be represented by a potential $U(x,y,z)$ where $x$ and $y$ are coordinate axes parallel to the mean plane of the wall.  Any eigenfunction of the motionless ground state of the system would be a function of all the coordinates of all the particles.  Therefore the boosted Galilean state cannot be an eigenstate of the system for any non-zero value of the velocity. 

Further, Landau's argumentation does not take into account quantum fluctuations.  We will show that by including the effects of surface roughness - a feature of all physical walls - and the effect of quantum fluctuations, the boundary conditions become much more complicated.

Let us consider that the walls containing the superfluid system are rough.  Then, as mentioned above, the state of the flowing superfluid cannot be a boosted ground state: there must be some changes in what Landau calls the inner motions close to the walls at any value of the speed, if only because disallowing a Galilean boost means the loss of a simple solution.  However, although the state resulting from a Galilean boost cannot be an eigenstate, this does not necessarily mean that there is no eigenstate. As we shall see, the physical picture is quite complex.

One might attempt to adapt the smooth-wall argument to the rough-wall situation by including some boundary layer close to the wall, which would merge continuously with the boosted ground state far from the wall. This cannot be the case for several reasons.  The most obvious (but not direct) argument against this is the following: Think of a {\emph{classical}} gas flowing tangentially at a low speed (low Reynolds number) with respect to a plane wall; if this wall were perfectly smooth down to molecular scales and if the fluid particles were elastically scattered by this wall, there would be no exchange of longitudinal momentum between the wall and the classical flowing gas; but if the wall were rough, there would necessarily be friction because the perturbation brought to the velocity distribution in the gas is not localized in the neighborhood of the wall. The wall friction, at least in a permanent state, generates a constant shear stress on the fluid at any arbitrary distance from the wall of the finite system such as the long cylinder considered here (we are ignoring transient effects and assume the system has reached its steady state).  In this example, no boundary layer takes care of the roughness to allow merging with a state of uniform speed far from the wall. 

Returning to the superfluid case, a way of understanding what happens near a rough wall is to look at the dilute Bose gas limit, which is valid for recent experiments in ultracold trapped atomic gases \cite{pit, Pethick}.  While this limit is not strictly valid for superfluid liquid Helium, there has been a long history of its use to gain insight into superfluid Helium behavior.  In this limit an expansion in terms of the small parameter in the theory ($\sqrt{n_0 a^3}$, where $n_0$ is the number density and $a$ is the positive 2-body scattering length) allows one to apply perturbation theory. In this limit, the leading order is given by the time-independent Gross-Pitaevskii equation (GPE), a classical equation for the amplitude of the mean-field wavefunction $\Psi^{(0)}(\vec{r})$:
\begin{equation}
\label{GPE}
(\hat T + U(\vec{r}) - \mu) \Psi^{(0)}(\vec{r})+g | \Psi^{(0)}(\vec{r}) |^2 \Psi^{(0)}(\vec{r}) = 0,
\end{equation}
where $\mu$ is the chemical potential, $\hat T \equiv - \frac{\hbar^2 \nabla^2}{2 m}+i \hbar v_s \frac{\partial}{\partial x}+\frac{1}{2} m{v_s}^2$, $g=4 \pi \hbar^2 a /2 m$, and $m$ is the mass of the atoms.  A stationary potential in a moving flow would thus experience a force
\begin{equation}
{\vec F}_{GPE}=-\int d^3 \vec{r} |\hat \Psi^{(0)}(\vec{r})|^2 {\vec \nabla} U(\vec{r}).
\end{equation}
Eq. (\ref{GPE}) reduces in the hydrodynamic limit to the familiar Bernoulli equations of a compressible fluid (if one ignores the quantum-pressure term) if the maximum flow velocity does not exceed the speed of sound | the critical speed identified by Landau.  There is a well defined steady solution of the flow equation typically with a thin boundary layer compensating for the rough wall. If the maximum flow velocity exceeds Mach $1$, the steady solution of the GPE disappears and vortices are released \cite{pomvortex,mcann}. The behavior of the GPE agrees almost perfectly with Landau's conception of superfluid flow, where no irreversible losses occur below this threshold speed. However the GPE describes only part of the picture; it leaves out the zero-point quantum fluctuations.

\section{Effect of Quantum Fluctuations}
In the dilute limit, quantum fluctuations appear at the next order in the expansion parameter ($\sqrt{n_0 a^3}$).  (See Appendix for a review of the effect of scattering quantum fluctuations in the dilute limit and an outline of the calculation of the resulting drag force on a localized stationary object immersed in a superfluid moving at arbitrarily low velocities.)

Let us leave rough walls aside for a moment and focus on the effect of quantum fluctuations.  In a trivial sense, the fluid motion does not induce any transition in the occupation number of the quantum state.  But one has to define exactly what is meant by occupation number here and, more specifically, what state one is referring to. The ground state refers to the state that is immobile in the lab frame. In this state, the macroscopic wavefunction corresponds to a uniform state, and zero-point fluctuations (defined as the ground state of occupation for certain modes of fluctuations) also have a well defined amplitude.  Practically, each mode of fluctuation is associated with a certain wavenumber. Because of the global Galilean invariance of the system (ground state and fluctuations) any fluid motion can also be seen as changing the quantum state of the system by a Galilean boost.  This Galilean boost defines exactly the quantum state. 

If one assumes the wall to be flat down to molecular scales then, regardless of whether the fluid is a classical one at finite temperature or a superfluid at zero temperature, there would be no rubbing between the flowing fluid and the wall itself because no force is generated in a given direction if the potential does not vary in that direction.  Therefore any tangential stress generated from the scattering of quantum fluctuations by the boundary would have to come from a modulation of the potential wall normal to the flow direction.

We now reintroduce rough walls.  The eigenstates of this system where quantum fluctuations scatter off the rough walls differ from those of a homogeneous Galilean-boosted system of ground state plus excitations.  It is nevertheless instructive to map the true eigenstates of our system with rough walls onto those of the latter system since at large hydrodynamic scales the only non-entropy producing states are those of a Galilean-boosted homogeneous superfluid component plus excitations.  The ground eigenstate of our scattering system maps onto the Galilean-boosted ground state plus similarly Galilean-boosted excitations, i.e. the normal fluid component, because, in general, mapping a ground state of one system onto another system whose Hamiltonian has different boundary conditions will result in nonzero amplitudes in the excited states of eigenstates describing the target system.  

This mechanism for creating excitations differs from Landau's theory of macroscopic excitation of quasiparticles that does not depend on the scattering conditions of the rough walls and that requires speeds larger than Landau's critical speed.  The interaction between the rough wall potential and the quantum fluctuations is a scattering problem as described in \cite{roberts1, roberts2}, and the relative difference of speed between the superfluid and the wall changes the scattering condition.  (To discuss equilibrium, one must at least implicitly assume that the higher order terms where the excitations interact, i.e. the Landau and Beliaev damping rates (see \cite{pit, damping1,damping2,damping3,damping4,damping5} and references therein), are the source of collisions required for the system to equilibrate.)  The temperature difference between this normal fluid component and the boundary must depend on the tangential superfluid velocity and vanish if this velocity is zero. Below we present some ideas for representing this boundary condition. 

\section{Effective Boundary Condition}
\subsection{Leading order (assuming $v_s$ is constant)}
Taking into account the scattering of quantum fluctuations as described above, the simplest representation of the boundary condition at the interface between a superfluid and its containing wall begins with the assumption that the scattering of the quantum fluctuations (also boosted by this local speed) generates a temperature difference between the wall and the superfluid system, resulting in some normal fluid component at rest along the boundary even if the boundary is held at $T_b=0$.  This temperature difference is proportional to the square of $v_{s}$ | the tangential superflow speed | since $v_{s}$ and $(-v_{s})$ have the same effect.  Ultimately, in equilibrium there should be no ongoing generation of entropy (we expand upon this idea below).  To make a boundary condition out of this assumption one has to assume that, in the finite system, if the temperature difference is not that at which this equilibrium condition is reached, the tangential speed should induce some net transformation of superfluid component into normal fluid component or vice versa at the boundary.  (This is reminiscent of what occurs when there is a normal heat flux at the boundary, often referred to as Kapitza resistance \cite{kapitza1} although it is important to stress that we only consider a perfectly insulating boundary condition in this paper.)  One can then write a similar phenomenological non-equilibrium fluid mechanics boundary condition (valid for scales much larger than that of the surface roughness) for such a situation, yielding the main result of this paper: 
\begin{equation}
\vec{n} \cdot (\vec{j}_n - \vec{j}_s)=\alpha (T_b-T) +\beta v_s^2
\label{eq:firstattempt}
\end{equation}
where $\vec{n}$ is the unit vector normal to the average (macroscopic) surface, $\vec{j}_n$ and $\vec{j}_s$ are the mass fluxes for the normal fluid component and superfluid component respectively, $T_b$ is the temperature at the boundary, and $T$ is the local temperature of the fluid.  $\alpha$ and $\beta$ are phenomenological parameters which depend on the surface roughness.  We arrived at this boundary condition by speculating that the equilibrium temperature is proportional to the superfluid component speed $v_s^2$ multiplied by a parameter characterizing the surface roughness and by assuming that $v_s$ is approximately constant.  This tangential boundary condition is to be supplemented by the total normal mass flux condition for a solid wall
\begin{equation}
\label{sw}
\vec{n} \cdot (\vec{j}_n + \vec{j}_s) = 0
\end{equation}
as well as by the normal condition for energy transport across the boundary \cite{kha}, which is responsible for second sound but has no bearing on the present discussion as we do not consider energy flux across the boundary.  

At thermal equilibrium, $\vec{n} \cdot \vec{j}_n =0$ (actually $\vec{j}_n =0$) because there should be no continuous generation or absorption of entropy.  Therefore, $\vec{n} \cdot \vec{j}_s =0$ (from eq. (\ref{sw})) together with eq. (\ref{eq:firstattempt}) yield the following expression for equilibrium temperature $T$ of the fluid:
\begin{equation}
\label{temp}
T = T_{b} + \frac{\beta v_s^2}{\alpha}.
\end{equation}
Let us note here the peculiar thermal equilibrium in the situation considered in this paper.  If $v_s=0$ in the situation considered in this paper, thermal equilibrium would imply that the normal fluid component (and thus the superfluid system) would be at the same temperature as the boundary; but at a nonzero $v_s$ the scattering of quantum fluctuations will necessarily force a temperature difference between the boundary and superfluid system at equilibrium.  That is to say, $T$ would be greater than $T_b$ in thermal equilibrium.  This can occur because the usual argument for uniform $T$ at equilibrium as a result of energy exchange in thermodynamics does not apply here, since energy is also stored in the motion of the superfluid component.  The scattering of quantum fluctuations in a moving superfluid would cause conversion from superfluid component into normal fluid component and, with that, the kinetic energy of the superfluid component is converted into heat.  By the relation eq. (\ref{temp}), this means that $\frac{\beta}{\alpha}$ must be greater than zero.  Thus even if $T_b$ is held at zero and $v_s$ is constrained to be nonzero, $T$ will be nonzero.  

\subsection{Higher order correction (local boundary condition)}
Although to leading order $v_{s}$ can be treated as constant, the roughness of the surface would cause some slight variation in $v_s$ across a given flow structure.  Eq. (\ref{eq:firstattempt}) does not have the flexibility to permit equilibrium between the non-constant $v_s$ and a constant wall temperature, a constant temperature $T$ inside the fluid and no net transformation between normal and superfluid components at the wall (as represented by the left-hand side of equation (\ref{eq:firstattempt})).  

To permit a nonuniform $v_s$, we will consider a modification of eq. (\ref{eq:firstattempt}).  Recall first that in thermal equilibrium $T > T_b$ if $v_s$ is nonzero.  Because the system must be at thermal equilibrium, to take into account scattering by the rough walls one must assume that, apart from the coherent superfluid component, there is some amount of normal fluid component moving with the container.  This is a property of the system as a whole and cannot be written simply as a local boundary condition as was attempted in eq. (\ref{eq:firstattempt}). Nevertheless we do expect a local relation between the various physical parameters at a given point on the wall. To reconcile the existence of both a global thermal equilibrium state and a local boundary condition, we suggest that instead of a smooth function of $\alpha (T_b-T) +\beta v_s^2$, the RHS of eq. (\ref{eq:firstattempt}) is modulated by a Heaviside function, i.e. 
\begin{equation}
\label{heaviside}
\vec{n} \cdot (\vec{j}_n - \vec{j}_s)= (\alpha (T_b-T) +\beta v_s^2)\,H(-\alpha (T_b-T) -\beta v_s^2),
\end{equation} where $H(x) = 1$ for positive $x$ and zero otherwise. With this modification, the RHS of eq. (\ref{eq:firstattempt}) vanishes as soon as $\alpha (T_b-T) +\beta v_s^2$ becomes positive.  This rather unusual boundary condition seems to agree with all the constraints of the problem, but constitutes only a small correction to eq. (\ref{eq:firstattempt}).  Eq. (\ref{heaviside}) implies that, for a given constant wall temperature and a given field of wall speed $v_s$, equilibrium will be reached at a temperature $T$ in the fluid such that 
\begin{equation}
T = T_{b} + \frac{\beta}{\alpha} \mathrm{max}(v_s^2)
\end{equation} where $\mathrm{max}(v_s^2)$ is the system-wide maximum speed at the wall.  Note again that even if $T_b = 0$, the temperature of the fluid will go to a nonzero $T$ for a finite system and nonzero $v_s$.     


\section{Discussion and Consequences of Effective Boundary Condition}
It is important to understand the difference between this picture and the friction in Landau's theory.  Here, the friction with the wall is not an ongoing process, unlike the continuous shedding of vortices beyond the critical Landau speed; instead the `friction' in this picture (namely the scattering of quantum fluctuations) brings the system {\emph{reversibly}} to a state of equilibrium where there is some amount of normal fluid component at rest with respect to the wall. This fits well with the widely observed existence of persistent currents, something that would be clearly incompatible with friction by continuous scattering of quantum fluctuations. This analysis only applies to a closed geometry where multiple scattering occurs, like the moving container we alluded to before. Indeed all experiments are done in finite containers.  However, we expect that equilibrium will be hard to reach in geometries approaching, for example, the classical Stokes situation of a moving sphere in a fluid at rest at infinity. In such a case, equilibrium will be reached only when the temperature due to the fluid motion near the sphere equilibrates with the temperature `at infinity', at the wall of the container.  In a semi-infinite geometry - a superfluid flow over a single wall - all of the superfluid component will eventually be converted to the normal fluid component as equilibrium will never be reached. This can be likened to the classical fluid mechanism of a flow in a semi-infinite geometry | the boundary layer will grow forever.

There are some important consequences of the boundary condition postulated in the present article, eq. (\ref{eq:firstattempt}).  (The correction resulting from the introduction of the Heaviside function (eq. \ref{heaviside}) is slight and has no material effect on the conclusions drawn below.)  Firstly, if there is a superfluid flow along a rough boundary, there will always be a small amount of normal fluid component at rest with respect to the wall, even if $T_b$ is kept at 0.  This normal fluid component,  at least in principle, could be observed in (immersed) torsion pendulum experiments \cite{andronikashvili}.  It would appear as a nonlinear correction to the pendulum equation; at increased angular velocity more normal fluid is carried by the oscillating discs, so the effective moment of inertia depends on the angular velocity.  Secondly, this boundary condition provides a critical speed in bulk flow experiments, one above which the equilibrium temperature in the flow is larger than the thermodynamic temperature of transition.  This might provide an alternative to the problematic vortex-shedding hypothesis \cite{vortexshedding} used to explain the infamous discrepancy (two to four orders of magnitude difference) between the observed critical speed and that predicted by Landau. Thirdly, this boundary condition tells us that when one slows down the moving boundary, excitations are converted into superfluid component (much like superfluid component creation when heat is removed from the boundary in counterflow experiments).  This newly created superfluid (coherent) component is likely to be not truly coherent, much like a Bose-Einstein condensate growing out of a nonequilibrium gas, which has a phase coherence length growing continuously with time after its inception - typically as $\left(\frac{\hbar t}{m}\right)^ {1/2}$ from the time the condensate is created.  In the situation considered, $t$ is the approximate time it takes for the newly formed superfluid component to travel a distance $\xi$ from the wall.  A rough assumption of the phase coherence length, which is approximately the average distance between vortices, for our system can be had by substituting $t \sim \xi /u_s$ \cite{diff} for $t$, where $u_s$ is the component of the superfluid velocity normal to the boundary, i.e. $u_s = \vec{n} \cdot \vec{v_s}$ .  (This substitution requires the assumption that the long range order of the phase depends on vortices carried by the superfluid.)  Therefore the coherence length parallel to the wall is  $\left(\frac{\hbar \xi}{m u_s}\right)^{1/2}$.  Almost all vortices will have disappeared if this phase coherence length is larger than the size of the system, and vortices will be everywhere when this length is much smaller than the length of the system, with crossover length scale for the size of the system being $\frac{m u_s}{\hbar}$.  Because $\frac{m}{\hbar}$ has the physical dimensions of a shear viscosity, this condition can be likened to a condition on a Reynolds number.

Although the boundary condition was motivated by superfluid Helium experiments, the exquisite control over the atomic interactions and confining potential in experiments with current dilute Bose-Einstein condensates make this an ideal medium to test systematically the predictions presented in this paper and shed light on superfluid Helium flow experiments past a rough wall.  In addition to previously mentioned atomic chip experiments where surface roughness is important, a possible example would be to drag a laser sheet, where the surface roughness can be precisely controlled, through a condensate and measure the resulting heating of the gas.  Similarily, one could modify a version of a previous experiment where a laser was dragged through a condensate \cite{ketterle}: one should be able to give the laser beam roughness using an acoustic-optical modulator and twist the beam in the condensate so that it acts like a rough corrugated cylinder rotating in a dilute BEC.  Another possible experiment would be a superfluid flow over a disordered optical lattice \cite{oplattice1,oplattice2,oplattice3} or optical speckle potential \cite{speckle} which could be adjusted and mimic the effects of surface roughness.  Finally, experiments to observe persistent currents in dilute BECs are just beginning to be realized \cite{pc1,pc2}, and one could, in principle, precisely modify the trap in such a way as to control the surface roughness, unlike the experiments on superfluid Helium, and detect the normal fluid component of the gas by measuring the temperature and mass flux.  

D.C.R. gratefully acknowledge stimulating discussions with Brian Anderson and Kevin Henderson.

\appendix
\section{Scattering of quantum fluctuations}

In the dilute limit, quantum fluctuations appear at the next order in the expansion parameter ($\sqrt{n_0 a^3}$) defined by expanding the exact bosonic field operator in terms of a coherent field plus a small quantum fluctuating operator, i.e. 
\begin{equation}
\hat \psi({\vec{r}},t) = \Psi^{(1)}({\vec{r}},t) +\hat \phi({\vec{r}},t),
\end{equation}  
where $\Psi^{(1)}$ is the wavefunction of the condensate affected by the action of the quantum fluctuations.  A stationary potential in a moving superfluid in this dilute limit would then experience a force 
\begin{equation}
{\vec F}=-\int d^3 {\vec{r}} (| \Psi^{(1)}(\vec{r})|^2+ \langle \hat \phi^\dag (\vec{r}) \hat \phi (\vec{r}) \rangle_{T=0}){\vec \nabla} U(\vec{r}).
\end{equation}
$\hat \phi$ can be expressed as the sum over excited states of quasiparticle operators $\alpha_k$ and $\alpha_k^\dag$ weighted by quasiparticle amplitudes $u_k(\vec{r})$ and $v_k(\vec{r})$ such that
\begin{equation}
\label{qp} 
\hat \phi(\vec{r})=\sum_{k'} \left( u_{k}(\vec{r}) \hat \alpha_{k} -v^*_{k}(\vec{r}) \hat \alpha_{k}^\dag \right). 
\end{equation}
$u_k(\vec{r})$ and $v_k(\vec{r})$ are assumed to satisfy the dynamic equations that, in this dilute limit, govern the quantum fluctuations: the Bogoliubov-de Gennes equations (for derivations and explanations see for example \cite{pit,fetter1,fetter2}), 
\begin{equation}
\label{BdG1} \hat {\cal L} u_{k}(\vec{r}) -g(\Psi^{(0)})^2 v_{k}(\vec{r}) = E_k u_{ k}(\vec{r})
\end{equation}
\begin{equation}
\label{BdG2} \hat {\cal L}^*  v_{k}(\vec{r})- g(\Psi^{(0)*})^2 u_{k}(\vec{r}) = -E_k v_{ k}(\vec{r}),
\end{equation}
where $E_k$ is the eigenvalue associated with momentum state $\vec{k}$, $\hat {\cal L} = \hat T + U(\vec{r}) - \mu +2 g|\Psi^{(0)}|^2$, and $^*$ signifies complex conjugate.  The weakly interacting Hamiltonian would then be diagonalized by $\alpha_k$ and $\alpha_k^\dag$.
In this limit, the behavior of the coherent field modified by the quantum fluctuations is given by the Generalized GPE \cite{ggpe1,ggpe2}
\begin{equation}
 (\hat T+U(\vec{r})-\mu) \Psi^{(1)}(\vec{r})+g | \Psi^{(1)}(\vec{r}) |^2 \Psi^{(1)}(\vec{r})
+ \sum_{k'} 2 g |v_k(\vec{r})|^2\Psi^{(1)}(\vec{r})-g \sum_{k'}  u_k(\vec{r})v^*_k(\vec{r})\Psi^{(1)}(\vec{r})-f(\vec{r})\Psi^{(1)}(\vec{r})=0.
\end{equation}
The term $g \sum_{k'}  u_k(\vec{r})v^*_k(\vec{r})\Psi^{(1)}(\vec{r})$ requires renormalization as the contact potential approximation causes it to be ultraviolet divergent. In the final term on the LHS, $f(r)=\sum_{k'} c_k v^*_k(\vec{r}),$ where $c_k=g \int d^3{\vec{r}}  | \Psi^{(1)}(\vec{r}) |^2(\Psi^{(1)*}(\vec{r}) u_k(\vec{r})+ \Psi^{(1)}(\vec{r}) v_k(\vec{r}))$.  \hspace{0.5 cm} $f(\vec{r}) \Psi^{(1)}(\vec{r})$ ensures that the excited modes and the condensate are orthogonal to each other.  \cite{roberts1, roberts2} provide explicit solutions of these equations describing an effective scattering problem for the quantum fluctuations around a stationary object in a superfluid moving at any arbitrarily small velocity, and demonstrate the existence of a nonzero drag force on the object in such a situation.  It is important to note that the situation they consider is an infinitely extended system, where the kind of thermal equilibrium described in this article cannot be reached.


\begin{thebibliography}{9}
\bibitem{vortexshedding} 
The usual explanation for this gap between predicted and observed critical velocity involves hypothesizing that vortex shedding takes place at low speeds. This explanation is challenged by two facts.  First, solutions of the Gross-Pitaevskii equation (GPE) only exhibit vortex shedding at speeds on the same order as the speed of sound \cite{pomvortex}, which is the critical Landau speed for dilute Bose-Einstein condensates. Second, in order to reach a metastable vortex state, a nucleation barrier must be crossed; the Arrhenius factor for the creation of vortices large enough to lower the critical speed significantly makes this barrier crossing extremely unlikely, with probabilities on the order of a miniscule $e^{-10^8}$ \cite{putt}.  In fact, it has been shown \cite{pomeaurica} within the GPE context that the onset of vortex shedding in a superfluid flow is at the velocity where an unstable steady vortex state and a stable vortex-less state merge at a saddle-node bifurcation beyond which no steady flow exists.  This supports the claim that a high energy barrier must be crossed in order for vortices to be emitted.  The height of the nucleation barrier is associated with the energy of the (unstable) steady vortex state. Therefore the vortex-shedding hypothesis has many problems explaning observed critical speed in Helium in a physically consistent way.
\bibitem{freeslip} This lack of friction with the walls is usually represented in viscous fluids by the free-slip boundary condition ${\vec {n}} \cdot \frac {\partial u}{\partial {\vec{r}}}=0$ \cite{koplik} where $u$ is the tangential speed and $\vec{n}$ the normal to the wall.  However, it is well known that the free-slip boundary condition does not preclude dissipation in the bulk of a viscous fluid flow, except for exactly parallel flows (i.e. not to rough-boundary situations).
\bibitem{kapitza} While this paper deals only with flow tangential to the wall, there has been interest in a transverse energy transport boundary condition related to Kapitza resistance.  See for example \cite{kapitza1} and \cite{pollack}.
\bibitem{heatflux} It is worth noting that, unlike ordinary fluids, there is a strong coupling in superfluids between heat flux and fluid motion, because heat (or entropy) can be carried by normal fluid only.  Therefore, a heat flux at a solid boundary can induce a superfluid flow toward the boundary where it is then transformed into an outgoing flow of normal fluid.
\end{thebibliography}
\end{document}